# MIMO Z-Interference Channels: Capacity Under Strong and Noisy Interference

Xiaohu Shang
Department of EE
Princeton University
Email: xshang@princeton.edu

Biao Chen
Department of EECS
Syracuse University
Email: bichen@syr.edu

Gerhard Kramer
Department of EE
University of Southern California
Email: gkramer@usc.edu

H. Vincent Poor
Department of EE
Princeton University
Email: poor@princeton.edu

*Abstract*—The capacity regions of multiple-input multiple-output Gaussian Z-interference channels are established for the very strong interference and aligned strong interference cases. The sum-rate capacity of such channels is established under noisy interference. These results generalize known results for scalar Gaussian Z-interference channels.

## I. INTRODUCTION

Determining the capacity of a multiple-input multiple-output (MIMO) interference channel (IC) is a long standing open problem. Recently, by assuming that some channel matrices are invertible, [1] derived the capacity region of MIMO ICs with average power constraints under strong interference, and sum-rate capacities under noisy interference and mixed interference. Those results were extended to MIMO ICs with average covariance constraints in [2]. Later, the noisy interference sum-rate capacity of a MIMO IC with an average power constraint was also considered in [3]. While the corresponding result in [1] requires that all the input covariance matrices satisfy a closed-form condition, [3] requires that the optimal solution of a non-convex optimization problem be non-singular and satisfy a complex condition. We note that neither [1] nor [3] includes the other as a special case.

In this paper, we consider the capacity of a MIMO Z-interference channel (ZIC) in which the received signals are defined as

$$\boldsymbol{y}_1 = \mathbf{H}_1\boldsymbol{x}_1 + \mathbf{F}\boldsymbol{x}_2 + \boldsymbol{z}_1 \quad \text{and}$$
$$\boldsymbol{y}_2 = \mathbf{H}_2\boldsymbol{x}_2 + \boldsymbol{z}_2, \qquad (1)$$

where $\boldsymbol{x}_i$ is the transmitted signal of user $i$, $i = 1, 2$; $\mathbf{H}_i$ and $\mathbf{F}$ are channel matrices known at both transmitters and receivers; and $\boldsymbol{z}_i$ is a zero-mean circularly symmetric complex Gaussian random vector with identity covariance matrix, i.e., $\boldsymbol{z}_i \sim \mathcal{CN}(\mathbf{0}, \mathbf{I})$. Transmitter $i$ and receiver $i$ have $t_i$ and $r_i$ antennas, respectively. The transmitted signal $\boldsymbol{x}_i$ is subject to a power constraint, denoted as $\mathcal{P}$, that takes a form from any one of the following:

$$\sum_{j=1}^{n} E\left[\boldsymbol{x}_{ij}\boldsymbol{x}_{ij}^\dagger\right] \preceq n\bar{\mathbf{S}}_i, \qquad (2)$$

This work was supported in part by the National Science Foundation under Grants CNS-09-05398, CCF-09-05320 and CCF-09-05235.

$$\sum_{j=1}^{n} \text{tr}\left(E\left[\boldsymbol{x}_{ij}\boldsymbol{x}_{ij}^\dagger\right]\right) \leq n\bar{P}_i, \qquad (3)$$

$$\text{tr}\left(E\left[\boldsymbol{x}_{ij}\boldsymbol{x}_{ij}^\dagger\right]\right) \leq P_i, \quad j = 1, \cdots, n, \quad \text{or} \qquad (4)$$

$$\sum_{j=1}^{n} \left(E\left[\boldsymbol{x}_{ij}\boldsymbol{x}_{ij}^\dagger\right]\right)_k \leq n\bar{P}_{ik}, \quad k = 1, \cdots, t_i, \qquad (5)$$

where the channel is assumed to be used $n$ times and $\boldsymbol{x}_{ij}$ is the transmitted signal of user $i$ at time $j$. Here, $E[\cdot]$ denotes expectation; $(\cdot)^\dagger$ denotes the Hermitian of a matrix; $\mathbf{A} \succeq \mathbf{B}$ means that $\mathbf{A}$ and $\mathbf{B}$ are both semi-positive definite Hermitian matrices and $\mathbf{A} - \mathbf{B}$ is also semi-positive definite; $\text{tr}(\cdot)$ denotes the trace of a matrix; and $(\cdot)_k$ denotes the $k$th diagonal element of a square matrix. Constraints (2)-(5) are referred to respectively as the *expected block covariance* constraint, the *expected block power* constraint, the *expected per-symbol power* constraint, and the *expected per-antenna block power* constraint.

In this paper, we generalize the results of [1] and [2] to the cases in which the channel matrices can be arbitrary, and the constraint can be any one from (2)-(5). Specifically, we derive the capacity regions under very strong and aligned strong interference which are achieved by fully decoding the interference; and the sum-rate capacity under noisy interference which is achieved by treating the interference as noise.

The rest of the paper is organized as follows: our main results are summarized in Section II and the proofs are given in Section III; numerical examples are provided in Section IV and we conclude in Section V.

## II. MAIN RESULTS

In this section, we give the capacity regions for MIMO ZICs under very strong interference and aligned strong interference, and the sum-rate capacity under noisy interference.

*Theorem 1:* For the MIMO ZIC defined in (1) under any constraint $\mathcal{P}$ in (2)-(5), if

$$\log\left|\mathbf{I} + \mathbf{H}_1\mathbf{S}_1^*\mathbf{H}_1^\dagger + \mathbf{F}\mathbf{S}_2^*\mathbf{F}^\dagger\right| \geq \sum_{i=1}^{2} \log\left|\mathbf{I} + \mathbf{H}_i\mathbf{S}_i^*\mathbf{H}_i^\dagger\right|, (6)$$

where $|\cdot|$ denotes the determinant and

$$\mathbf{S}_i^* = \arg\max_{\mathbf{S}_i} \log\left|\mathbf{I} + \mathbf{H}_i\mathbf{S}_i\mathbf{H}_i^\dagger\right|, \quad i = 1, 2, \qquad (7)$$

then the capacity region is

$$\left\{ R_1 \leq \log \left| \mathbf{I} + \mathbf{H}_1 \mathbf{S}_1^* \mathbf{H}_1^\dagger \right|, R_2 \leq \log \left| \mathbf{I} + \mathbf{H}_2 \mathbf{S}_2^* \mathbf{H}_2^\dagger \right| \right\}. \quad (8)$$

We say that a MIMO ZIC has *very strong interference* if (6) is satisfied. In this case the interference does not reduce the capacity region. When both users transmit at the maximum rate, receiver 1 can first decode the interference by treating the desired signal as noise, i.e., we have

$$I\left(\boldsymbol{x}_2^*; \boldsymbol{y}_1^*\right) \geq I\left(\boldsymbol{x}_2^*; \boldsymbol{y}_2^* \mid \boldsymbol{x}_1^*\right),$$

where $\boldsymbol{x}_i^* \sim \mathcal{CN}\left(\mathbf{0}, \mathbf{S}_i^*\right)$ and $\boldsymbol{y}_i^*$ is defined in (1) with $\boldsymbol{x}_i$ replaced by $\boldsymbol{x}_i^*$, $i=1,2$. As with the scalar Gaussian IC where the notion of very strong interference depends on both the channel coefficients and the power constraints, for the MIMO IC our definition of very strong interference involves both the channel matrices and the constraint specified by $\mathcal{P}$. On setting $\mathbf{H}_i = 1$, $\mathbf{F} = \sqrt{a}$ and $\mathbf{S}_i = P_i$, then (6) becomes $a \geq 1 + P_1$. Therefore, Theorem 1 generalizes the capacity region for scalar Gaussian ICs under very strong interference [4].

*Theorem 2:* For the MIMO IC defined in (1) under any constraint $\mathcal{P}$ in (2)-(5), if there exists a matrix $\mathbf{A}$ such that

$$\mathbf{H}_2 = \mathbf{AF} \quad (9)$$
$$\text{and} \quad \mathbf{I} \succeq \mathbf{AA}^\dagger, \quad (10)$$

then the capacity region is

$$\bigcup_{(\mathbf{S}_1, \mathbf{S}_2) \in \mathcal{P}} \left\{ \begin{array}{c} R_1 \leq \log \left| \mathbf{I} + \mathbf{H}_1 \mathbf{S}_1 \mathbf{H}_1^\dagger \right| \\ R_2 \leq \log \left| \mathbf{I} + \mathbf{H}_2 \mathbf{S}_2 \mathbf{H}_2^\dagger \right| \\ R_1 + R_2 \leq \log \left| \mathbf{I} + \mathbf{H}_1 \mathbf{S}_1 \mathbf{H}_1^\dagger + \mathbf{FS}_2 \mathbf{F}^\dagger \right| \end{array} \right\}. \quad (11)$$

Theorem 2 gives the capacity region of a MIMO ZIC under *aligned strong interference*. Equation (9) means that $\mathbf{H}_2$ is a linear transformation of $\mathbf{F}$. Therefore, users 1 and 2 see $\boldsymbol{x}_2$ in the forms of $\mathbf{F}\boldsymbol{x}_2$ and $\mathbf{AF}\boldsymbol{x}_2$, respectively. If $\mathbf{A}^\dagger \mathbf{A} \preceq \mathbf{I}$, then user 1 can decode $\boldsymbol{x}_2$ if user 2 can.

We can also verify Theorem 2 in a way similar to that done in [5] and [6] for scalar Gaussian ICs under strong interference. Assuming the rate pair $(R_1, R_2)$ is achievable, then $\boldsymbol{x}_1$ and $\boldsymbol{x}_2$ can be reliably recovered at user 1 and user 2, respectively. After subtracting $\boldsymbol{x}_1$ from $\boldsymbol{y}_1$, user 1 obtains

$$\boldsymbol{y}_1' = \mathbf{F}\boldsymbol{x}_2 + \boldsymbol{z}_1. \quad (12)$$

We can pre-multiply $\boldsymbol{y}_1'$ by $\mathbf{A}$ and get

$$\boldsymbol{y}_1'' = \mathbf{AF}\boldsymbol{x}_2 + \mathbf{A}\boldsymbol{z}_1$$
$$= \mathbf{H}_2 \boldsymbol{x}_2 + \mathbf{A}\boldsymbol{z}_1. \quad (13)$$

If $\mathbf{A}^\dagger \mathbf{A} \preceq \mathbf{I}$ then $\mathbf{AA}^\dagger \preceq \mathbf{I}$ and the received signal at user 2 can be written as

$$\boldsymbol{y}_2 = \mathbf{H}_2 \boldsymbol{x}_2 + \boldsymbol{z}_2$$
$$= \boldsymbol{y}_1'' + \boldsymbol{w}, \quad (14)$$

where $\boldsymbol{w} \sim \mathcal{CN}\left(\mathbf{0}, \mathbf{I} - \mathbf{AA}^\dagger\right)$, and $\boldsymbol{w}$ is independent of $\boldsymbol{y}_1''$. Since $\boldsymbol{x}_2$ can be recovered from $\boldsymbol{y}_2$, $\boldsymbol{x}_2$ can also be recovered from $\boldsymbol{y}_1''$. Thus, user 1 can decode both $\boldsymbol{x}_1$ and $\boldsymbol{x}_2$. The above development imposes no structure on $\boldsymbol{x}_i$, $i=1,2$. Therefore, as long as $\boldsymbol{x}_2$ (which can be non-Gaussian with arbitrary covariance matrix) can be decoded by receiver 2, it can also be decoded by receiver 1.

When $\mathbf{F}$ is left invertible, it can be shown that (9) and (10) are equivalent to $\mathbf{F}^\dagger \mathbf{F} \succeq \mathbf{H}_2^\dagger \mathbf{H}_2$. Therefore, Theorem 2 includes as its special cases, the capacity regions of both single-input multiple-output (SIMO) ICs with per-symbol power constraints under strong interference [7] and scalar ICs with block power constraints under strong interference [5], [6].

Remark 1: Conditions (9) and (10) are sufficient conditions for a MIMO ZIC to have aligned strong interference under any of the constraints in (2)-(5). However, these conditions can be relaxed for a specified constraint. For example, [2] shows that under the expected block covariance constraint (2), if there exist matrices $\mathbf{A}$ and $\mathbf{B}$ that satisfy

$$\mathbf{H}_2 = \mathbf{AF} + \mathbf{B},$$
$$\mathbf{I} \succeq \mathbf{A}^\dagger \mathbf{A}$$
$$\text{and} \quad \bar{\mathbf{S}}_2 \mathbf{B}^\dagger = \mathbf{0}, \quad (15)$$

then this MIMO ZIC has aligned strong interference and the capacity is achieved by letting receiver 1 decode both $\boldsymbol{x}_1$ and $\boldsymbol{x}_2$.

Remark 2: In contrast to the situation for scalar ICs, very strong interference for a MIMO ZIC is no longer a special case of aligned strong interference, i.e., there exist channels that satisfy (6) but not (9) and (10). An example is given in Section IV.

Remark 3: Adapting the definition of a two-sided discrete memoryless IC under strong interference in [8] to a one-sided IC, we require

$$I\left(\boldsymbol{x}_2; \boldsymbol{y}_2 \mid \boldsymbol{x}_1\right) \leq I\left(\boldsymbol{x}_2; \boldsymbol{y}_1 \mid \boldsymbol{x}_1\right) \quad (16)$$

to be satisfied for all possible distributions of $\boldsymbol{x}_1$ and $\boldsymbol{x}_2$. We now show that conditions (9) and (10) imply (16) for all possible distributions of $\boldsymbol{x}_1$ and $\boldsymbol{x}_2$. We have

$$\begin{aligned} I\left(\boldsymbol{x}_2; \boldsymbol{y}_2 \mid \boldsymbol{x}_1\right) &= I\left(\boldsymbol{x}_2; \mathbf{H}_2 \boldsymbol{x}_2 + \boldsymbol{z}_2\right) \\ &= I\left(\boldsymbol{x}_2; \mathbf{AF}\boldsymbol{x}_2 + \boldsymbol{z}_2\right) \\ &\stackrel{(a)}{=} I\left(\boldsymbol{x}_2; \mathbf{A}\left(\mathbf{F}\boldsymbol{x}_2 + \boldsymbol{z}_1\right) + \boldsymbol{w}\right) \\ &\stackrel{(b)}{\leq} I\left(\boldsymbol{x}_2; \mathbf{A}\left(\mathbf{F}\boldsymbol{x}_2 + \boldsymbol{z}_1\right)\right) \\ &\stackrel{(c)}{\leq} I\left(\boldsymbol{x}_2; \mathbf{F}\boldsymbol{x}_2 + \boldsymbol{z}_1\right), \end{aligned} \quad (17)$$

where in (a) we define $\boldsymbol{w} \sim \mathcal{CN}\left(\mathbf{0}, \mathbf{I} - \mathbf{AA}^\dagger\right)$, and $\boldsymbol{w}$ is independent of $\boldsymbol{x}_2$ and $\boldsymbol{z}_1$; and hence we have (b); and (c) follows from the Markov relationship $\boldsymbol{x}_2 \to \mathbf{F}\boldsymbol{x}_2 + \boldsymbol{z}_1 \to \mathbf{A}\left(\mathbf{F}\boldsymbol{x}_2 + \boldsymbol{z}_1\right)$.

*Theorem 3:* For the MIMO IC defined in (1) under any constraint $\mathcal{P}$ in (2)-(5), if there exists a matrix $\mathbf{A}$ such that

$$\mathbf{F} = \mathbf{A}^\dagger \mathbf{H}_2 \quad (18)$$

and $\quad \mathbf{I} \succeq \mathbf{A}\mathbf{A}^\dagger,$ (19)

then the sum-rate capacity is

$$\max_{(\mathbf{S}_1,\mathbf{S}_2)\in\mathcal{P}} \left[\log\left|\mathbf{I}+\mathbf{H}_1\mathbf{S}_1\mathbf{H}_1^\dagger\left(\mathbf{I}+\mathbf{F}\mathbf{S}_2\mathbf{F}^\dagger\right)^{-1}\right| + \log\left|\mathbf{I}+\mathbf{H}_2\mathbf{S}_2\mathbf{H}_2^\dagger\right|\right]. \quad (20)$$

Theorem 3 gives the noisy-interference sum-rate capacity of a MIMO ZIC. Specifically, when (18) and (19) are satisfied, the sum-rate capacity can be achieved by treating interference as noise. When $\mathbf{H}_2$ is left invertible, conditions (18) and (19) are equivalent to $\mathbf{F}^\dagger\mathbf{F} \preceq \mathbf{H}_2^\dagger\mathbf{H}_2$. Therefore, Theorem 3 includes the scalar Gaussian ZIC noisy-interference sum-rate capacity [9], [10] as a special case[1].

Remark 4: As with Theorem 2, Theorem 3 gives a sufficient condition for a MIMO ZIC to have noisy interference under any constraint in (2)-(5). If the constraint is specified, then this condition can be relaxed. For example, [2] shows that under the expected block covariance constraint (2), if there exist matrices $\mathbf{A}$ and $\mathbf{B}$ that satisfy

$$\mathbf{F} = \mathbf{A}^\dagger\mathbf{H}_2 + \mathbf{B},$$
$$\mathbf{I} \succeq \mathbf{A}\mathbf{A}^\dagger$$
and $\quad \bar{\mathbf{S}}_2\mathbf{B}^\dagger = \mathbf{0},$ (21)

then this MIMO ZIC has noisy interference.

We now present a theorem that generalizes Theorem 3.

*Theorem 4:* For the MIMO ZIC defined in (1) under any constraint $\mathcal{P}$ in (2)-(5), if the optimal solution $\mathbf{S}_1^*, \mathbf{S}_2^*$ and $\mathbf{A}^*$ for

$$\min_{\mathbf{A}} \max_{(\mathbf{S}_1,\mathbf{S}_2)} \left[\log\left|\mathbf{I}+\mathbf{H}_1\mathbf{S}_1\mathbf{H}_1^\dagger+\mathbf{F}\mathbf{S}_2\mathbf{F}^\dagger\right|\right.$$
$$+ \log\left|\mathbf{I}+\mathbf{H}_2\mathbf{S}_2\mathbf{H}_2^\dagger - (\mathbf{H}_2\mathbf{S}_2\mathbf{F}^\dagger+\mathbf{A})\right.$$
$$\left.\left.\cdot\left(\mathbf{I}+\mathbf{F}\mathbf{S}_2\mathbf{F}^\dagger\right)^{-1}(\mathbf{H}_2\mathbf{S}_2\mathbf{F}^\dagger+\mathbf{A})^\dagger\right|\right]$$
subject to $\quad (\mathbf{S}_1,\mathbf{S}_2)\in\mathcal{P},$
$$\mathbf{A}\mathbf{A}^\dagger \preceq \mathbf{I}, \quad (22)$$

satisfies

$$\mathbf{S}_2^*\mathbf{F}^\dagger = \mathbf{S}_2^*\mathbf{H}_2^\dagger\mathbf{A}^*, \quad (23)$$

then this MIMO ZIC has noisy interference and the sum-rate capacity is (20).

If (18) is satisfied, (23) is also satisfied, and then the optimal value of (22) can be shown to be (20). Therefore, Theorem 3 is a special case of Theorem 4.

## III. PROOFS OF THE MAIN RESULTS

### A. Preliminaries

We first introduce some lemmas which will be used in our proof.

---

[1]The case with $a < 1$ is often referred to as ZIC with weak interference in the literature. We use the term noisy-interference simply because of the fact that treating interference as noise achieves the sum-rate capacity.

*Lemma 1:* [2, Lemma 2] Let $\boldsymbol{x}^k = \{\boldsymbol{x}_1,\cdots,\boldsymbol{x}_k\}$ and $\boldsymbol{y}^k = \{\boldsymbol{y}_1,\cdots,\boldsymbol{y}_k\}$ be two sequences of random vectors, and let $\boldsymbol{x}^*$ and $\boldsymbol{y}^*$ be Gaussian vectors with covariance matrices satisfying

$$\mathrm{Cov}\begin{bmatrix}\boldsymbol{x}^*\\\boldsymbol{y}^*\end{bmatrix} = \frac{1}{k}\sum_{i=1}^k \mathrm{Cov}\begin{bmatrix}\boldsymbol{x}_i\\\boldsymbol{y}_i\end{bmatrix},$$

then we have

$$h\left(\boldsymbol{x}^k\right) \leq k\cdot h\left(\widehat{\boldsymbol{x}}^*\right)$$
and $\quad h\left(\boldsymbol{y}^k\,|\,\boldsymbol{x}^k\right) \leq k\cdot h\left(\widehat{\boldsymbol{y}}^*\,|\,\widehat{\boldsymbol{x}}^*\right).$

*Lemma 2:* [2, Lemma 5] Let $\boldsymbol{x}, \boldsymbol{u}$ and $\boldsymbol{v}$ be jointly Gaussian vectors, such that $\boldsymbol{x}$ is independent of $\boldsymbol{u}$ and $\boldsymbol{v}$. Denote $\mathrm{Cov}(\boldsymbol{x}) = \mathbf{S}_x$, $\mathrm{Cov}(\boldsymbol{u}) = \mathbf{S}_u$ and $\mathrm{Cov}(\boldsymbol{u},\boldsymbol{v}) = \mathbf{S}_{uv}$. If $\mathbf{S}_u$ is invertible, then $\boldsymbol{x} \to \mathbf{H}\boldsymbol{x}+\boldsymbol{u} \to \mathbf{G}\boldsymbol{x}+\boldsymbol{v}$ forms a Markov chain if and only if

$$\mathbf{S}_x\mathbf{G}^\dagger = \mathbf{S}_x\mathbf{H}^\dagger\mathbf{S}_u^{-1}\mathbf{S}_{uv}.$$

### B. Proof of Theorem 1

The converse follows by giving receiver 1 the message not intended for it and applying the maximum-entropy theory to show that Gaussian input distributions are optimal. To prove achievability, let $\boldsymbol{x}_i \sim \mathcal{CN}(\mathbf{0},\mathbf{S}_i^*)$, $i=1,2$, and let user 1 transmit at rate $R_1 = \log\left|\mathbf{I}+\mathbf{H}_1\mathbf{S}_1^*\mathbf{H}_1^\dagger\right|$, and user 2 transmit at rate $R_2 = \log\left|\mathbf{I}+\mathbf{H}_2\mathbf{S}_2^*\mathbf{H}_2^\dagger\right|$. Inequality (6) guarantees that user 1 can first decode $\boldsymbol{x}_2$ by treating $\boldsymbol{x}_1$ as noise. After the interference is subtracted, user 1 sees a single-user Gaussian MIMO channel. Therefore, the rate region (8) is achievable.

### C. Proof of Theorem 2

Suppose the channel is used $n$ times. The transmitted and received vector sequences are denoted by $\boldsymbol{x}_i^n$ and $\boldsymbol{y}_i^n$ for user $i$, $i=1,2$, and $\boldsymbol{x}_i^n$ satisfies $\mathcal{P}$. We further let $\bar{\boldsymbol{x}}_i^* \sim \mathcal{CN}(\mathbf{0},\mathbf{S}_i)$, where

$$\mathbf{S}_i = \frac{1}{n}\sum_{j=1}^n \mathrm{Cov}(\boldsymbol{x}_{ij}). \quad (24)$$

Since $\mathbf{A}^\dagger\mathbf{A} \preceq \mathbf{I}$, there exists a Gaussian random vector $\boldsymbol{n}$ whose joint distribution with $\boldsymbol{z}_2$ is

$$\begin{bmatrix}\boldsymbol{z}_2\\\boldsymbol{n}\end{bmatrix} \sim \mathcal{CN}\left(\mathbf{0},\begin{bmatrix}\mathbf{I} & \mathbf{A}\\\mathbf{A}^\dagger & \mathbf{I}\end{bmatrix}\right). \quad (25)$$

Moreover, from (9), $\boldsymbol{n}$ is of the same dimension as $\boldsymbol{z}_1$ and hence has the same marginal distribution as $\boldsymbol{z}_1$.

Let $\epsilon > 0$ and $\epsilon \to 0$ as $n\to+\infty$. From Fano's inequality, any achievable rates must satisfy

$n(R_1+R_2) - n\epsilon$
$\leq I(\boldsymbol{x}_1^n;\boldsymbol{y}_1^n) + I(\boldsymbol{x}_2^n;\boldsymbol{y}_2^n)$
$\leq I(\boldsymbol{x}_1^n;\boldsymbol{y}_1^n) + I(\boldsymbol{x}_2^n;\boldsymbol{y}_2^n,\mathbf{F}\boldsymbol{x}_2^n+\boldsymbol{n}^n)$
$= h(\mathbf{H}_1\boldsymbol{x}_1^n+\mathbf{F}\boldsymbol{x}_2^n+\boldsymbol{z}_1^n) - h(\mathbf{F}\boldsymbol{x}_2^n+\boldsymbol{z}_1^n) - h(\boldsymbol{z}_2^n\,|\,\boldsymbol{n}^n)$
$\quad + h(\mathbf{F}\boldsymbol{x}_2^n+\boldsymbol{n}^n) - h(\boldsymbol{n}^n) + h(\mathbf{H}_2\boldsymbol{x}_2^n+\boldsymbol{z}_2^n\,|\,\mathbf{F}\boldsymbol{x}_2^n+\boldsymbol{n}^n)$

$$\overset{(a)}{=} I(\boldsymbol{x}_1^n, \boldsymbol{x}_2^n; \mathbf{H}_1\boldsymbol{x}_1^n + \mathbf{F}\boldsymbol{x}_2^n + \boldsymbol{z}_1^n) - h(\boldsymbol{z}_2^n \mid \boldsymbol{n}^n)$$
$$+ h(\mathbf{H}_2\boldsymbol{x}_2^n + \boldsymbol{z}_2^n \mid \mathbf{F}\boldsymbol{x}_2^n + \boldsymbol{n}^n)$$
$$\overset{(b)}{\leq} I(\boldsymbol{x}_1^n, \boldsymbol{x}_2^n; \mathbf{H}_1\boldsymbol{x}_1^n + \mathbf{F}\boldsymbol{x}_2^n + \boldsymbol{z}_1^n) - nh(\boldsymbol{z}_2 \mid \boldsymbol{n})$$
$$+ nh(\mathbf{H}_2\bar{\boldsymbol{x}}_2^* + \boldsymbol{z}_2 \mid \mathbf{F}\bar{\boldsymbol{x}}_2^* + \boldsymbol{n})$$
$$\overset{(c)}{=} I(\boldsymbol{x}_1^n, \boldsymbol{x}_2^n; \mathbf{H}_1\boldsymbol{x}_1^n + \mathbf{F}\boldsymbol{x}_2^n + \boldsymbol{z}_1^n) - nh(\boldsymbol{z}_2 \mid \boldsymbol{n})$$
$$+ nh(\mathbf{H}_2\bar{\boldsymbol{x}}_2^* + \boldsymbol{z}_2 \mid \mathbf{F}\bar{\boldsymbol{x}}_2^* + \boldsymbol{n}, \bar{\boldsymbol{x}}_2^*)$$
$$= I(\boldsymbol{x}_1^n, \boldsymbol{x}_2^n; \mathbf{H}_1\boldsymbol{x}_1^n + \mathbf{F}\boldsymbol{x}_2^n + \boldsymbol{z}_1^n)$$
$$\leq n \log \left| \mathbf{I} + \mathbf{H}_1 \mathbf{S}_1 \mathbf{H}_1^\dagger + \mathbf{F} \mathbf{S}_2 \mathbf{F}^\dagger \right|, \tag{26}$$

where $\boldsymbol{z}_i^n = \left[ \boldsymbol{z}_{i,1}^\dagger, \boldsymbol{z}_{i,2}^\dagger, \ldots, \boldsymbol{z}_{i,n}^\dagger \right]^\dagger$ and $\boldsymbol{n}^n = \left[ \boldsymbol{n}_1^\dagger, \boldsymbol{n}_2^\dagger, \ldots, \boldsymbol{n}_n^\dagger \right]^\dagger$, $i = 1, 2$, and $\left[ \boldsymbol{z}_{2,j}^\dagger, \boldsymbol{n}_j^\dagger \right]^\dagger$, $j = 1, \ldots, n$, are independent and identically distributed (i.i.d.) as (25).

Equality (a) is from the fact that $\boldsymbol{n}$ and $\boldsymbol{z}_1$ have the same marginal distribution. Inequality (b) is by Lemma 1. $\bar{\boldsymbol{x}}_1^*$ is independent of $\bar{\boldsymbol{x}}_2^*$ and $\bar{\boldsymbol{y}}_i^*$ is defined in (1) with $\boldsymbol{x}_i$ replaced by $\bar{\boldsymbol{x}}_i^*$. Equality (c) is from (9) which means
$$\mathbf{S}_2 \mathbf{H}_2^\dagger = \mathbf{S}_2 \mathbf{F}^\dagger \mathbf{A}^\dagger.$$

By Lemma 2, $\bar{\boldsymbol{x}}_2^* \to \mathbf{F}\bar{\boldsymbol{x}}_2^* + \boldsymbol{n} \to \mathbf{H}_2\bar{\boldsymbol{x}}_2^* + \boldsymbol{z}_2$ forms a Markov chain.

Therefore, (11) is an outer bound for the capacity region. On the other hand, (11) is also achievable by requiring user 1 to decode messages from both users. This completes the proof of Theorem 2.

*D. Proof of Theorem 3*

We define $\boldsymbol{n}$, $\bar{\boldsymbol{x}}_1^*$ and $\bar{\boldsymbol{x}}_2^*$ as in the proof of Theorem 2. From Fano's inequality, any achievable rates must satisfy

$$n(R_1 + R_2) - n\epsilon$$
$$\leq I(\boldsymbol{x}_1^n; \boldsymbol{y}_1^n) + I(\boldsymbol{x}_2^n; \boldsymbol{y}_2^n)$$
$$\leq I(\boldsymbol{x}_1^n; \boldsymbol{y}_1^n) + I(\boldsymbol{x}_2^n; \boldsymbol{y}_2^n, \mathbf{F}\boldsymbol{x}_2^n + \boldsymbol{n}^n)$$
$$= h(\mathbf{H}_1\boldsymbol{x}_1^n + \mathbf{F}\boldsymbol{x}_2^n + \boldsymbol{z}_1^n) - h(\mathbf{F}\boldsymbol{x}_2^n + \boldsymbol{z}_1^n) - h(\boldsymbol{n}^n)$$
$$+ h(\mathbf{F}\boldsymbol{x}_2^n + \boldsymbol{n}^n) + h(\mathbf{H}_2\boldsymbol{x}_2^n + \boldsymbol{z}_2^n \mid \mathbf{F}\boldsymbol{x}_2^n + \boldsymbol{n}^n)$$
$$- h(\boldsymbol{z}_2^n \mid \boldsymbol{n}^n)$$
$$\overset{(a)}{=} h(\mathbf{H}_1\boldsymbol{x}_1^n + \mathbf{F}\boldsymbol{x}_2^n + \boldsymbol{z}_1^n) - h(\boldsymbol{n}^n)$$
$$+ h(\mathbf{H}_2\boldsymbol{x}_2^n + \boldsymbol{z}_2^n \mid \mathbf{F}\boldsymbol{x}_2^n + \boldsymbol{n}^n) - h(\boldsymbol{z}_2^n \mid \boldsymbol{n}^n)$$
$$\overset{(b)}{\leq} nh(\mathbf{H}_1\bar{\boldsymbol{x}}_1^* + \mathbf{F}\bar{\boldsymbol{x}}_2^* + \boldsymbol{z}_1) - nh(\boldsymbol{n})$$
$$+ nh(\mathbf{H}_2\bar{\boldsymbol{x}}_2^* + \boldsymbol{z}_2 \mid \mathbf{F}\bar{\boldsymbol{x}}_2^* + \boldsymbol{n}) - nh(\boldsymbol{z}_2 \mid \boldsymbol{n}) \tag{27}$$
$$= nh(\mathbf{H}_1\bar{\boldsymbol{x}}_1^* + \mathbf{F}\bar{\boldsymbol{x}}_2^* + \boldsymbol{z}_1) + nh(\mathbf{H}_2\bar{\boldsymbol{x}}_2^* + \boldsymbol{z}_2) - nh(\boldsymbol{n})$$
$$+ nh(\mathbf{F}\bar{\boldsymbol{x}}_2^* + \boldsymbol{n} \mid \mathbf{H}_2\bar{\boldsymbol{x}}_2^* + \boldsymbol{z}_2) - nh(\mathbf{F}\bar{\boldsymbol{x}}_2^* + \boldsymbol{n})$$
$$- nh(\boldsymbol{z}_2 \mid \boldsymbol{n})$$
$$\overset{(c)}{=} nh(\mathbf{H}_1\bar{\boldsymbol{x}}_1^* + \mathbf{F}\bar{\boldsymbol{x}}_2^* + \boldsymbol{z}_1) + nh(\mathbf{H}_2\bar{\boldsymbol{x}}_2^* + \boldsymbol{z}_2) - nh(\boldsymbol{n})$$
$$+ nh(\mathbf{F}\bar{\boldsymbol{x}}_2^* + \boldsymbol{n} \mid \mathbf{H}_2\bar{\boldsymbol{x}}_2^* + \boldsymbol{z}_2, \bar{\boldsymbol{x}}_2^*) - nh(\mathbf{F}\bar{\boldsymbol{x}}_2^* + \boldsymbol{n})$$
$$- nh(\boldsymbol{z}_2 \mid \boldsymbol{n})$$
$$\overset{(d)}{=} nh(\mathbf{H}_1\bar{\boldsymbol{x}}_1^* + \mathbf{F}\bar{\boldsymbol{x}}_2^* + \boldsymbol{z}_1) - nh(\boldsymbol{n}) + nh(\mathbf{H}_2\bar{\boldsymbol{x}}_2^* + \boldsymbol{z}_2)$$
$$+ nh(\boldsymbol{n} \mid \boldsymbol{z}_2) - nh(\mathbf{F}\bar{\boldsymbol{x}}_2^* + \boldsymbol{z}_1) - nh(\boldsymbol{z}_2 \mid \boldsymbol{n})$$

$$= nh(\mathbf{H}_1\bar{\boldsymbol{x}}_1^* + \mathbf{F}\bar{\boldsymbol{x}}_2^* + \boldsymbol{z}_1) - nh(\mathbf{F}\bar{\boldsymbol{x}}_2^* + \boldsymbol{z}_1)$$
$$+ nh(\mathbf{H}_2\bar{\boldsymbol{x}}_2^* + \boldsymbol{z}_2) - nh(\boldsymbol{z}_2)$$
$$= n \log \left| \mathbf{I} + \mathbf{H}_1 \mathbf{S}_1 \mathbf{H}_1^\dagger \left( \mathbf{I} + \mathbf{F}\mathbf{S}_2 \mathbf{F}^\dagger \right)^{-1} \right|$$
$$+ n \log \left| \mathbf{I} + \mathbf{H}_2 \mathbf{S}_2 \mathbf{H}_2^\dagger \right|, \tag{28}$$

where (a) and (d) follow because $\boldsymbol{n}$ and $\boldsymbol{z}_1$ have the same marginal distribution; (b) is from Lemma 1; and (c) is from (18) which means
$$\mathbf{S}_2 \mathbf{F}^\dagger = \mathbf{S}_2 \mathbf{H}_2^\dagger \mathbf{A}.$$

By Lemma 2, $\bar{\boldsymbol{x}}_2^* \to \mathbf{H}_2\bar{\boldsymbol{x}}_2^* + \boldsymbol{z}_2 \to \mathbf{F}\bar{\boldsymbol{x}}_2^* + \boldsymbol{n}$ forms a Markov chain.

Since (20) is achievable, the sum-rate capacity is (20) if (18) and (19) hold. Therefore, Theorem 3 is proved.

*E. Proof of Theorem 4*

Achievability is straightforward, so we only need to show the converse. We again define $\bar{\boldsymbol{x}}_1^*$ and $\bar{\boldsymbol{x}}_2^*$ as in the proof of Theorem 2, but $\boldsymbol{n}$ is defined as
$$\begin{bmatrix} \boldsymbol{z}_2 \\ \boldsymbol{n} \end{bmatrix} \sim \mathcal{CN}\left(\mathbf{0}, \begin{bmatrix} \mathbf{I} & \mathbf{A}^* \\ \mathbf{A}^{*\dagger} & \mathbf{I} \end{bmatrix}\right). \tag{29}$$

Following the proof of Theorem 3, from (27) we have

$$n(R_1 + R_2) - n\epsilon$$
$$\leq nh(\mathbf{H}_1\bar{\boldsymbol{x}}_1^* + \mathbf{F}\bar{\boldsymbol{x}}_2^* + \boldsymbol{z}_1) - nh(\boldsymbol{n})$$
$$+ nh(\mathbf{H}_2\bar{\boldsymbol{x}}_2^* + \boldsymbol{z}_2 \mid \mathbf{F}\bar{\boldsymbol{x}}_2^* + \boldsymbol{n}) - nh(\boldsymbol{z}_2 \mid \boldsymbol{n})$$
$$= n\log\left|\mathbf{I} + \mathbf{H}_1\mathbf{S}_1\mathbf{H}_1^\dagger + \mathbf{F}\mathbf{S}_2\mathbf{F}^\dagger\right| + n\log\left|\mathbf{I} + \mathbf{H}_2\mathbf{S}_2\mathbf{H}_2^\dagger - \right.$$
$$\left. \left(\mathbf{H}_2\mathbf{S}_2\mathbf{F}^\dagger + \mathbf{A}^*\right)\left(\mathbf{I} + \mathbf{F}\mathbf{S}_2\mathbf{F}^\dagger\right)^{-1}\left(\mathbf{H}_2\mathbf{S}_2\mathbf{F}^\dagger + \mathbf{A}^*\right)^\dagger\right|$$
$$\overset{(a)}{\leq} n\log\left|\mathbf{I} + \mathbf{H}_1\mathbf{S}_1^*\mathbf{H}_1^\dagger + \mathbf{F}\mathbf{S}_2^*\mathbf{F}^\dagger\right| + n\log\left|\mathbf{I} + \mathbf{H}_2\mathbf{S}_2^*\mathbf{H}_2^\dagger - \right.$$
$$\left. \left(\mathbf{H}_2\mathbf{S}_2^*\mathbf{F}^\dagger + \mathbf{A}^*\right)\left(\mathbf{I} + \mathbf{F}\mathbf{S}_2^*\mathbf{F}^\dagger\right)^{-1}\left(\mathbf{H}_2\mathbf{S}_2^*\mathbf{F}^\dagger + \mathbf{A}^*\right)^\dagger\right|$$
$$= nh(\mathbf{H}_1\boldsymbol{x}_1^* + \mathbf{F}\boldsymbol{x}_2^* + \boldsymbol{z}_1) - nh(\boldsymbol{z}_1)$$
$$+ nh(\mathbf{H}_2\boldsymbol{x}_2^* + \boldsymbol{z}_2 \mid \mathbf{F}\boldsymbol{x}_2^* + \boldsymbol{n}) - nh(\boldsymbol{z}_2 \mid \boldsymbol{n})$$
$$\overset{(b)}{=} nh(\mathbf{H}_1\boldsymbol{x}_1^* + \mathbf{F}\boldsymbol{x}_2^* + \boldsymbol{z}_1) - nh(\mathbf{F}\boldsymbol{x}_2^* + \boldsymbol{z}_1)$$
$$+ nh(\mathbf{H}_2\boldsymbol{x}_2^* + \boldsymbol{z}_2) - nh(\boldsymbol{z}_2)$$
$$= n\log\left|\mathbf{I} + \mathbf{H}_1\mathbf{S}_1^*\mathbf{H}_1^\dagger\left(\mathbf{I} + \mathbf{F}\mathbf{S}_2^*\mathbf{F}^\dagger\right)^{-1}\right|$$
$$+ n\log\left|\mathbf{I} + \mathbf{H}_2\mathbf{S}_2^*\mathbf{H}_2^\dagger\right|, \tag{30}$$

where (a) is from the fact that $(\mathbf{S}_1^*, \mathbf{S}_2^*)$ is optimal for (22); and in (b) we define $\boldsymbol{x}_i^* \sim \mathcal{CN}(\mathbf{0}, \mathbf{S}_i^*)$, $i = 1, 2$.

## IV. NUMERICAL EXAMPLES

*Example 1:* Consider a MIMO IC with expected block power constraint (3) and
$$\mathbf{H}_1 = \mathbf{H}_2 = \mathbf{I}, \quad \mathbf{F} = \begin{bmatrix} 1.0 & 0.5 \\ 0.8 & -1.8 \end{bmatrix} \text{ and } \bar{P}_1 = \bar{P}_2 = 2.$$

From (7) we have $\mathbf{S}_1^* = \mathbf{S}_2^* = \mathbf{I}$,
$$\log\left|\mathbf{I} + \mathbf{H}_1\mathbf{S}_1^*\mathbf{H}_1^\dagger + \mathbf{F}\mathbf{S}_2^*\mathbf{F}^\dagger\right| = 2.7408 \quad \text{and}$$

$$\log\left|\mathbf{I}+\mathbf{H}_1\mathbf{S}_1^*\mathbf{H}_1^\dagger\right|=\log\left|\mathbf{I}+\mathbf{H}_2\mathbf{S}_2^*\mathbf{H}_2^\dagger\right|=0.6931.$$

Then (6) is satisfied. Therefore, this MIMO ZIC has very strong interference and the capacity region is

$$\{(R_1,R_2):\quad 0\leq R_1\leq 0.6931,\quad 0\leq R_2\leq 0.6931\}.$$

However, the aligned strong interference conditions (9) and (10) for this channel are not satisfied, since $\mathbf{A}=\mathbf{F}^{-1}$ and $\mathbf{A}^\dagger\mathbf{A}\not\preceq\mathbf{I}$. Therefore, this MIMO ZIC has very strong interference but not aligned strong interference.

*Example 2:* Consider a MIMO IC with

$$\mathbf{H}_1=\mathbf{H}_2=\mathbf{I}\quad\text{and}\quad\mathbf{F}=\begin{bmatrix}0.3 & 0\\ 0 & 0.9\end{bmatrix}.$$

It is easy to see that this MIMO ZIC satisfies conditions (18) and (19), therefore it has noisy interference and the sum-rate capacity is achieved by treating interference as noise.

If this MIMO ZIC has expected block power constraint (3) and

$$\bar{P}_1=1\quad\text{and}\quad\bar{P}_2=4,$$

then the sum-rate capacity is $C=2.8138$ and the optimal input covariance matrices are[2]

$$\mathbf{S}_1^*=\begin{bmatrix}1 & 0\\ 0 & 0\end{bmatrix}\quad\text{and}\quad\mathbf{S}_2^*=\begin{bmatrix}1.84 & 0\\ 0 & 2.16\end{bmatrix}.$$

That is, transmitter 1 allocates no power to the second antenna due to the limited power and the greater interference that the second antenna experiences as compared to the first antenna.

If this MIMO ZIC has expected per-antenna block power constraint (5) and

$$\bar{P}_{11}=\bar{P}_{12}=0.5\quad\text{and}\quad\bar{P}_{21}=\bar{P}_{22}=2,$$

then the sum-rate capacity is $C=2.7252$ and the optimal input covariances are

$$\mathbf{S}_1^*=\begin{bmatrix}0.5 & 0\\ 0 & 0.5\end{bmatrix}\quad\text{and}\quad\mathbf{S}_2^*=\begin{bmatrix}2 & 0\\ 0 & 2\end{bmatrix}.$$

Obviously, in this example the block power constraint includes the per-antenna block power constraint as a special case. Therefore, the former constraint results in a larger sum-rate capacity than the latter constraint.

*Example 3:* Consider a MIMO ZIC under expected block covariance constraint (2), and

$$\mathbf{H}_1=\mathbf{I},\quad\mathbf{F}=\begin{bmatrix}1.3 & 1.1 & 1.4\\ 1.5 & -0.5 & 3.0\\ 0.9 & -0.36 & 1.5\end{bmatrix},$$

$$\mathbf{H}_2=\begin{bmatrix}1.0 & 2.0 & 0.5\\ 1.0 & 1.0 & 2\\ 0.5 & 0.4 & 0.5\end{bmatrix},$$

$$\bar{\mathbf{S}}_1=\mathbf{I}\quad\text{and}\quad\bar{\mathbf{S}}_2=\begin{bmatrix}1.8 & 1.0 & -0.4\\ 1.0 & 5.0 & 2.0\\ -0.4 & 2.0 & 1.2\end{bmatrix}.$$

---

[2]The same results can be obtained from [11] since this MIMO ZIC is a parallel IC under noisy interference

From Theorem 3, we have $\mathbf{A}^\dagger=\mathbf{F}\mathbf{H}_2^{-1}$ and $\mathbf{A}\mathbf{A}^\dagger\not\preceq\mathbf{I}$. Therefore, Theorem 3 does not apply. However, from Theorem 4, the optimal solution of problem (22) is

$$\mathbf{A}=\begin{bmatrix}0.8 & 0 & 0\\ 0 & 0.5 & 0\\ 0 & 0 & 0.6\end{bmatrix},\mathbf{S}_1^*=\bar{\mathbf{S}}_1\text{ and }\mathbf{S}_2^*=\bar{\mathbf{S}}_2,$$

and we have

$$\mathbf{S}_2^*\mathbf{F}^\dagger=\mathbf{S}_2^*\mathbf{H}_2^\dagger\mathbf{A}^*$$

Therefore, this MIMO ZIC still has noisy interference and the sum-rate capacity $C=5.6622$ is achieved by treating interference as noise[3].

## V. Conclusion

The capacity regions of MIMO ZICs under very strong interference and aligned strong interference, and the sum-rate capacity under noisy interference have been obtained. The capacity results apply to various power constraints, and they extend the results of [1] and [2].

---

[3]The same result can be obtained from [2].